\documentclass[aps,prl,twocolumn,showpacs,groupedaddress]{revtex4}
\usepackage{amsmath}
\usepackage{amsfonts}
\usepackage{graphicx}
\usepackage{bm}
\usepackage{color}

\newcommand{\ket}[1]{\ensuremath{\left|{#1}\right\rangle}}

\newcommand{\oper}[1]{\hat{#1}}

\newcommand{\HG}{{\ensuremath{{\mathrm{HG}}}}}

\begin{document}

\title{Quantum communication without alignment using multiple-qubit single-photon states}

\author{L. Aolita}
\affiliation{Instituto de F\'{\i}sica, Universidade Federal do Rio de
Janeiro, Caixa Postal 68528, Rio de Janeiro, RJ 21941-972, Brazil}
\author{S. P. Walborn}
\affiliation{Instituto de F\'{\i}sica, Universidade Federal do Rio de
Janeiro, Caixa Postal 68528, Rio de Janeiro, RJ 21941-972, Brazil}

\date{\today}
\begin{abstract}
We propose a scheme for encoding logical qubits in a subspace protected against collective rotations around the propagation axis using the polarization and transverse
spatial degrees of freedom of single photons. This encoding allows for quantum key distribution without the need of a shared reference frame. We present methods
to generate entangled states of two logical qubits using present day down-conversion sources and linear optics, and show that the application of these entangled logical
states to quantum information schemes allows for alignment-free tests of Bell's inequalities, quantum dense coding and quantum teleportation.
\end{abstract}

\maketitle

The most common implementations of quantum communication schemes involve two or more parties that, in
order to encode and decode information, must share a common spatial
reference frame.
Nevertheless, it has been
pointed out that a shared reference frame (SRF) is a resource that should not be taken for granted, since establishing a perfect SRF requires the transmission of an infinite amount of
information \cite{peres01}. One can circumvent the need for a SRF encoding logical qubits in multi-qubit states with appropriate symmetry properties, so that the
states are rotationally invariant.  This results in a considerable reduction in overhead due to initial alignment stages; but, since all the available protocols exploiting multi--qubit  states of photons
require the use of two \cite{Wallton04, boileau0, chen06}, three \cite{boileau}, or four photons \cite{boileau, cabello03, bartlett03,fiorentino04} to encode one single logical qubit, this
increases the amount of resources as well as the sensitivity of the protocol to photon losses.
\par
The lack of alignment between two users of a protocol is equivalent to a collective random rotation of the qubits during the transmission process, which can be considered as a special
type of collective noise. Rotationally invariant states, in turn, span a decoherence-free subspace (DFS) protected against such noise.
A DFS protected against collective noise is a subspace of the total Hilbert space of a system that is immune to  decoherence, provided that it acts identically and simultaneously on each
member of the system \cite{palma96}.
 For example,
 two ions of the same species closely spaced in a Paul trap, or photons propagating close together in the same optical fiber, are exposed approximately to the same fluctuations.  Thus, through the use of DFSs,  their respective coherence properties can be enhanced considerably \cite{Kielpinsky01,chen06}. 
Nevertheless, the assumption of collective noise is in practice fulfilled only approximately, as two different particles are never actually
subject to exactly the same noise.
\par Photons are a natural candidate for quantum communication due to the ease with which they can be transmitted.  Using spontaneous
parametric down-conversion (SPDC), one can create triggered single photons or entangled photon pairs \cite{kwiat97,barreiro05}.
Also, there has been a great deal of recent work exploiting the fact that one can encode multiple qubits into multiple degrees of 
freedom (DOF) of photons \cite{barreiro05,oliveira05}. But as different DOF are not necessarily affected in the same way by the same rotation, the collective rotation hypothesis
is not necessarily valid, even for different DOF of the same single photon.  
\par
However, there do exist two DOF of the photon that play a
preferential role for alignment-free quantum
communication: the lack of a common reference frame in the
plane orthogonal to the propagation direction does imply a collective
rotation 
for the polarization and the transverse spatial DOF.
Here we show that it is possible to encode logical qubits into single photons using these two DOF, which allows
for the implementation of quantum information protocols in free space
without the need of aligning the two directions orthogonal to the
propagation axis. The novel aspect here is that the
rotationally invariant states are carried by single photons,
which not only reduces the required resources, but also the damage due to photon losses. We first present the encoding
scheme and then discuss the implementation of several quantum
communication protocols.
\par Let us outline the scenario.  Two users, $A$ and $B$, wish to communicate photonic qubits.  We assume that, in order to send and detect the photons, they have previously established a common propagation direction by, for example, using an intense laser beam. Our scheme is based on the fact that, in the usual paraxial approximation, both the polarization and transverse spatial modes of a field are defined in the plane transverse to the propagation direction of
the field.  For example, consider a photon prepared in the state $\ket{\psi_{A}} = \alpha\ket{H_A}+\beta\ket{V_A}$, where $\ket{H_A}$ and $\ket{V_A}$ refer
to the horizontal and vertical polarization directions in reference frame $A$.  A second user $B$ whose coordinate system is rotated an angle $\theta$ around the propagation
 axis, would describe this state as $\ket{\psi_B}=\alpha(\cos\theta \ket{H_B}+\sin\theta\ket{V_B})+ \beta(\cos\theta \ket{V_B}-\sin\theta\ket{H_B})$.
\par
\begin{figure}
 \includegraphics[width=6cm]{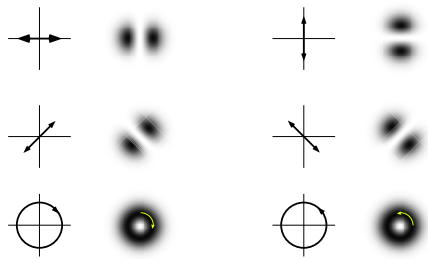}
 \caption{\label{fig:1} Polarization and HG modes of the electromagnetic field. Top: horizontal polarization $\ket{H}$ and $\HG_{01}$ mode
 $\ket{h}$ (left), vertical polarization $\ket{V}$ and $\HG_{10}$ mode $\ket{v}$ (right). Center: diagonal polarizations $(\ket{H} \pm \ket{V})/\sqrt{2}$ and diagonal modes $(\ket{h} \pm
 \ket{v})/\sqrt{2}$.  Bottom: circular polarizations $(\ket{H} \pm i\ket{V})/\sqrt{2}$ and circular modes $(\ket{h} \pm i\ket{v})/\sqrt{2}$.}
 \end{figure}

Similarly, one can encode information in the transverse spatial DOF of a photon using the Hermite Gaussian (HG) modes, denoted HG$_{nm}$, where the positive integers $n$ and $m$ are
the horizontal and vertical indices with respect to some coordinate system.  Here we consider the first order modes
($n+m=1$) HG$_{01}$ and HG$_{10}$, which form a basis analogous to that of polarization \cite{oneil00}, and are described relative
to its same reference
frame, as FIG. \ref{fig:1} shows. 
That is 
if a user $A$ can prepare an arbitrary quantum state $\ket{\Psi_A} = \alpha\ket{h_A}+\beta\ket{v_A}$,
where $h_A$ ($v_B$) represents a horizontally (vertically) aligned mode HG$_{10}$ (HG$_{01}$), then a  $\theta$-rotated user $B$ would describe this state as
$\ket{\Psi_B}=\alpha(\cos\theta \ket{h_B}+\sin\theta\ket{v_B})+ \beta(\cos\theta \ket{v_B}-\sin\theta\ket{h_B})$.  Since the effect of a rotated user is the same for
both DOF, the assumption of collective rotation, around the propagation axis,  is fulfilled exactly; and it is thus possible to construct rotationally invariant single photon states which form a basis
for a logical qubit.
 \par We define our single-photon logical computational basis $B_L$ using the same abstract encoding of \cite{Kielpinsky01, Wallton04}: $B_L=\{\ket{0_L}\equiv (\ket{Hv}-\ket{Vh})/\sqrt{2}, \ \ket{1_L}\equiv (\ket{Hh}+\ket{Vv})/\sqrt{2}\}$,
where the subindex $L$ stands for ``logical". Here lowercase (uppercase) letters refer to HG (polarization) modes; \emph{e.g.} $\ket{Hv}$ stands for a photon
with polarization $H$ and transverse mode $v$, etc. All states in the subspace $\mathbb{V}_{DFS}$, generated by $B_L$,
are invariant under arbitrary rotations of the reference frame around the propagation axis. These states are single-photon Bell states, and are easily 
prepared and detected unambiguously with perfect
efficiency using single photon controlled-not ($c\hat{\sigma}^{x}$) gates, which have been constructed using relatively simple linear optical devices \cite{fiorentino04}, and subsequently measuring in the physical basis using polarizing beam splitters (PBS) and transverse mode sorters (MS)\cite{sasada03}.
\par
Also one can manipulate the physical qubits individually using a combination of half-
and quarter-wave plates, in the case of polarization, or Dove prisms and mode converters, in the case of HG modes \cite{beijersbergen93,oneil00}.  Using these elements and
the interferometric techniques described in Refs. \cite{fiorentino04}, one can implement any controlled-logic operation among the physical qubits, and
consequently,  implement any $SU(2)$ unitary operation on the logical qubit.  For example, using a half-wave plate (HWP) aligned at $\phi/2$, followed by a HWP
aligned at $0^\circ$, realizes the polarization rotation $\hat{R}^{y}_{p}(2\phi)$: $\ket{H}\longrightarrow \cos\phi \ket{H} + \sin\phi \ket{V}$, $\ket{V}\longrightarrow -\sin\phi \ket{H} + \cos\phi \ket{V}$.
Under this physical transformation the logical states evolve like
$\ket{0_L} \rightarrow \cos\phi\ket{0_L} + \sin\phi \ket{1_L}$ and $\ket{1_L} \rightarrow  -\sin\phi \ket{0_L} + \cos\phi\ket{1_L}$, which corresponds to the logical
rotation $\hat{R}^{y}_{L}(2\phi)\equiv\exp(i\phi \hat{\sigma}^{y}_L)$, where $\hat{\sigma}^{y}_L$ is the usual Pauli operator in the logical basis.
It suffices now to show how to implement a logical rotation around any other axis. It is straightforward to see that, when acting
on  $\mathbb{V}_{DFS}$, the following identity holds:
$\hat{R}^{z}_{L}(\theta)\equiv c_{s}\hat{\sigma}^{x}_{p} . \hat{R}^{x}_{s}(-\theta) . c_{s}\hat{\sigma}^{x}_{p}$, where $p$ and $s$ refer to the
polarization and spatial mode qubits, respectively. For example, $c_{s}\hat{\sigma}^{x}_{p}$ is a controlled-not gate, where the polarization qubit is controlled by the spatial mode qubit. Finally, it is important to notice that, in contrast to $\hat{R}^{y}_{L}(2\phi)$, where the evolution takes place entirely
inside $\mathbb{V}_{DFS}$, in the case of $\hat{R}^{z}_{L}(\theta)$ the evolution is not fault-tolerant, meaning that $\hat{R}^{z}_{L}(\theta)$ takes the protected states
momentarily out of $\mathbb{V}_{DFS}$ and then brings them back. Nevertheless, this is of no consequence here, since  
$\hat{R}^{z}_{L}(\theta)$ is never applied during the transmission but at one of the user's laboratories, where the reference frame is perfectly defined.
\paragraph{Quantum key distribution.}
One immediate application of the ideas developed above is the implementation of single-photon quantum key distribution schemes which do not require
initial alignment of the preparation and measurement systems.  As an example, let us briefly outline the BB84 \cite{bb84} protocol using these single--photon logical qubits.  By choosing randomly between the angles
$\phi_A = \{0,\pi/4\}$ and applying the rotation $\hat{R}^{y}_{L}(2\phi_A)$, Alice can send random bits in either the logical computational basis or the rotated logical basis
$\ket{\pm_L}=(\ket{0_L}\pm\ket{1_L})/\sqrt{2}$.  Bob's random measurements in the computational or rotated logical bases, in turn, are carried out by simply
performing the logical rotations $\phi_B=\{0,-\pi/4\}$ and subsequently detecting --unambigously and with perfect efficiency--
in the computational logical basis $B_L$ with the single-photon Bell-state measurement (BSM) technique mentioned above.
\paragraph{Generation of entangled logical states.}
Using SPDC, it is possible to create two-photon states entangled in both polarization and transverse HG modes \cite{langford03,walborn05b}.  Pumping a single type-I
crystal with a Gaussian profile beam and post-selecting only the first-order modes, one can generate down-converted photons in the entangled state
$(\ket{Hh_{1}Hh_{2}}+\ket{Hv_{1}Hv_{2}})/\sqrt{2}$, where the subindices $1$ and $2$ refer to photons in different (longitudinal) momentum modes.  Using controlled
operations on the physical qubits, it is possible to transform this state to an
entangled logical state.  It is easy to show that applying the following sequence of gates $c_{s}\hat{\sigma}^{x}_{p} . \oper{h}_{s} . c_{s}\hat{\sigma}^{x}_{p}$ to
the physical qubits of each photon
transforms the SPDC output state to the two-photon Bell state composed of single-photon logical qubits:
$\ket{\Phi^+_L}=(\ket{0_{L_1}0_{L_2}}+\ket{1_{L_1}1_{L_2}})/\sqrt{2}$. Here $\oper{h}_{s}$ is a Hadamard gate on the spatial mode qubit. The other three logical Bell
states,  $\ket{\Phi^-_L}=(\ket{0_{L_1}0_{L_2}}-\ket{1_{L_1}1_{L_2}})/\sqrt{2}$ and $\ket{\Psi^\pm_L}=(\ket{0_{L_1}1_{L_2}}\pm\ket{1_{L_1}0_{L_2}})/\sqrt{2}$, can be
immediately obtained from $\ket{\Phi^+_L}$ with the single logical qubit rotations described before. An alternative method to generate an entangled logical state employs the
methods proposed in Ref. \cite{walborn05b} using a second-order HG pump beam. We note that this method can also be used to create non-maximally entangled logical states.
\paragraph{Tests of quantum nonlocality.} An important application of the entangled logical states just presented is the implemention of tests of
Bell's theorem without the usual alignment of the analyzers, with only two photons.
Furthermore, using non-maximally entangled logical states, one can implement alignment-free tests of Hardy's nonlocality without inequalities \cite{hardy93}.
The importance of such tests has been pointed in \cite{cabello03}, where a scheme using eight photons was proposed. Also, the ability to realize these tests allows for alignment-free implementations of entanglement-based quantum cryptography  \cite{ekert91}.
\paragraph{Bell state measurement.}    
\begin{figure}
\includegraphics[width=6cm]{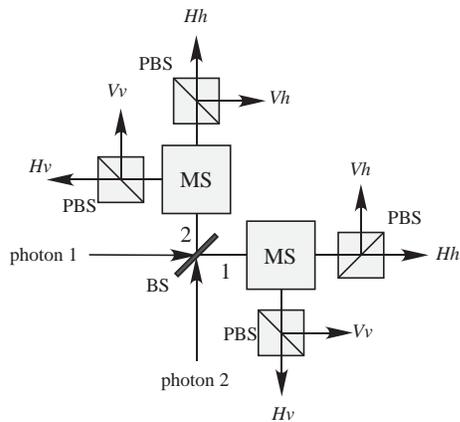}
\caption{\label{fig3} Bell-state measurement device for logical qubits using a 50-50 beam splitter (BS).  MS is a transverse mode sorter and PBS a polarizing beam
splitter.}
\end{figure}
The main ingredient to quantum teleportation and dense coding is a BSM on two qubits.  In previous experiments using qubits encoded in single DOF of photons, a
partial BSM was performed using two-photon interference \cite{mattle96,bouwmeester97}.  Here we present a device which can be used to perform a partial
BSM on the logical qubits presented here.  FIG. \ref{fig3} illustrates our analyzer. Photons are input on opposite sides of a 50-50 beam splitter (BS).  It has been shown that, in addition to the $\pi$ phase
shift present in the usual beam splitter transformations, the transverse spatial mode $\ket{h}$ acquires an additional $\pi$ phase due to reflection from the  BS
\cite{oliveira05}. That is, the evolution operator of a $50-50$ BS is $\hat{B}\equiv e^{i\frac{\pi}{2}(\hat{a}_{Hv}\hat{b}^{\dagger}_{Hv}+\hat{a}^{\dagger}_{Hv}\hat{b}_{Hv})}
\otimes e^{i\frac{\pi}{2}(\hat{a}_{Vv}\hat{b}^{\dagger}_{Vv}+\hat{a}^{\dagger}_{Vv}\hat{b}_{Vv})}\otimes
e^{-i\frac{\pi}{2}(\hat{a}_{Hh}\hat{b}^{\dagger}_{Hh}+\hat{a}^{\dagger}_{Hh}\hat{b}_{Hh})}\otimes
e^{-i\frac{\pi}{2}(\hat{a}_{Vh}\hat{b}^{\dagger}_{Vh}+\hat{a}^{\dagger}_{Vh}\hat{b}_{Vh})}$, where $\hat{a}^{\dagger}_{Xx}$ and $\hat{b}^{\dagger}_{Xx}$ are the creation
operators of one photon with polarization $X$ ($=H$ or $V$) and HG mode $x$ ($=h$ or $v$), in momentum mode $1$ and $2$, respectively (notice the different signs in
the exponents for the cases  $Xh$ and $Xv$). Upon application of $\hat{B}$ our logical states transform as:
\begin{eqnarray}
\label{eq:bellstatestransform}
\nonumber\ket{\Psi^{-}_L}\rightarrow\frac{1}{4}[(\ket{Hv_1 Hh_2}-\ket{Hh_1 Hv_2}-\ket{Hv_1 Vv_1}\\
\nonumber -\ket{Hv_2 Vv_2}
-\ket{Vh_1 Hh_1}-\ket{Vh_2 Hh_2}-\ket{Vh_1 Vv_2} \\
\nonumber +\ket{Vv_1 Vh_2})-(\ket{Hh_1 Hv_2} -\ket{Hv_1 Hh_2}\\
\nonumber
+\ket{Hh_1 Vh_1} +\ket{Hh_2 Vh_2} +\ket{Vv_1 Hv_1}\\
+\ket{Vv_2 Hv_2}-\ket{Vv_1 Vh_2}+\ket{Vh_1 Vv_2})]\ , \\
 \ket{\Psi^{+}_L}\rightarrow\frac{1}{4}[(\ket{Hv_1 Hh_1}-\ket{Hh_2 Hv_2}+\ket{Hv_1 Vv_2} \nonumber \\
\nonumber +\ket{Vv_1 Hv_2}
-\ket{Vh_1 Hh_2}-\ket{Hh_1 Vh_2}+\ket{Vh_1 Vv_1} \\
\nonumber -\ket{Vh_2 Vv_2})+(\ket{Hh_1 Hv_1}-\ket{Hh_2 Hv_2}\\
\nonumber
-\ket{Hh_1 Vh_2}-\ket{Vh_1 Hh_2}
+\ket{Vv_1 Hv_2}\\
+\ket{Hv_1 Vv_2}
 -\ket{Vv_2 Vh_2}+\ket{Vv_1 Vh_1})]\ ,  \\
\ket{\Phi^{\pm}_L}\rightarrow\frac{1}{4}[(\ket{2Hv_1}-\ket{2Hv_2}-\sqrt{2}\ket{Vh_1 Hv_2} \nonumber \\
\nonumber
-\sqrt{2}\ket{Hv_1 Vh_2}-\ket{2Vh_1} +\ket{2Vh_2} )
\\
\nonumber
\pm (\ket{2Hh_2}-\ket{2Hh_1} +\sqrt{2}\ket{Hh_1 Vv_2}\\
+\sqrt{2}\ket{Vv_1 Hh_2}+\ket{2Vv_1} -\ket{2Vv_2} )] \ .
\end{eqnarray}
Using additional linear optics devices to separate $H$ from $V$ and  $h$ from $v$, it is possible to separate
photons in different states.  For example, events like $\ket{Vh_1 Hh_1}$, corresponding to two photons in the same transverse mode and
 same output port of the BS, can still be separated by
polarization, and registered through two-photon coincidence detections. The states $\ket{\Psi^{-}_L}$ and $\ket{\Psi^{+}_L}$ always
give coincidences in different modes, so that they can be detected and discriminated with perfect efficiency. On the other hand, $\ket{\Phi^{-}_L}$ and
$\ket{\Phi^{+}_L}$ give coincidence events $50\,\%$
of the time \cite{comment0}. These
events are the same for both states, so they still cannot be discriminated from one another. Still, this BSM of logical qubits can be used for the
implementation of quantum dense coding, allowing for the transmission of $\log_{2}3$ bits of information in a single logical qubit, with an overall
efficiency of $1-\frac{1/3}{2}\approx 0.83$ with coincidence detections.
\paragraph{Quantum teleportation.}
 Quantum teleportation \cite{bennett93} is perhaps the most important application of a BSM. 
 For example, the error probability in the transmission of a qubit scales exponentially with the length of the channel, seriously limiting
quantum communications.  One way to circumvent this is using a quantum repeater  \cite{briegel98}, in which the channel between
$A$ and $B$ is divided into $N$ segments using $N$ entangled pairs. Since the procedure requires $N-1$ teleportation protocols,
$N$ SRF's are necessary. Therefore it is apparent that a SRF-free quantum teleportation scheme is of considerable importance, since it
would greatly reduce the overhead of the repeater.
\par Suppose that $A$ and $B$  share the entangled logical state $\ket{\Phi^+_{L}}_{12}$ between logical qubits $1$ and $2$, but do not a share common reference frame; and  $A$ would like
to teleport a third logical qubit $\ket{\psi_L}_3=\alpha\ket{0_L}_3+\beta\ket{1_L}_3$ to $B$.
Using the BSM presented above, $A$ can project logical qubits $1$ and $3$ onto the logical Bell basis, and can identify $\ket{\Psi^{\pm}_L}$ unambiguously with
100\% efficiency.   She then communicates her measurement result to $B$, who applies the logical operations $\hat{\sigma}^{x}_{L}$
or $\hat{\sigma}^{z}_{L}\hat{\sigma}^{x}_{L}$ to his
logical qubit when the measurement result is $\ket{\Psi^{+}_L}$ or $\ket{\Psi^{-}_L}$, respectively.  The  overall teleportation
efficiency is 50\%, and the teleportation fidelity  is $1$.
Nevertheless, $A$ and $B$ can increase the efficiency of the protocol, while still working in the coincidence basis, at the cost of a
slight decrease in fidelity. Half of the time, $A$'s BSM corresponds to one of the
$\ket{\Phi^{\pm}_L}$ states, which, in turn, give coincidence detections 50\% of the time.
In these cases, $A$ can inform $B$ that the measurement result was $\ket{\Phi^{\pm}_L}$, and then
$B$ knows that his logical qubit is either in state $\alpha\ket{0_L}+\beta\ket{1_L}$ (with probability $1/2$, corresponding to a teleportation fidelity of $1$) or in
$\alpha\ket{0_L}-\beta\ket{1_L}$ (probability $1/2$, corresponding to an average teleportation fidelity of $2/3$). Thus, if they perform the teleportation protocol
only when $A$  detects a coincidence event, the efficiency is improved to $75\%$ and the overall fidelity of the teleportation procedure is $2/3 \times 1 + 1/3 \times 1/2
\times 1 + 1/3 \times 1/2 \times 2/3 = 17/18 \approx 94.4$\%.
\paragraph{Discussion.}  In addition to rotations due to user misalignment, there is also the problem of decoherence of the physical
qubits.  The atmosphere is not birefringent, so random polarization rotations are not an issue in free space propagation of photons.
However, fluctuations of the refractive index of the atmosphere can deform the transverse modes.  Nonetheless, these effects could
be monitored using an intense reference beam, and then corrected.  Optical fibers preserve spatial mode but are birefringent,
which severely limits long-distance quantum communication with polarization qubits.  Depending on the communication protocol, these
polarization rotations can be overcome using a ``plug and play" setup \cite{zbinden97}.     
\begin{acknowledgments}
We thank L. Davidovich for reading the manuscript and acknowledge financial support from the CNPq, CAPES, FAPERJ, and the Brazilian Millennium Institute for Quantum Information.
\end{acknowledgments}


\begin{thebibliography}{99}
\bibitem{comment0}
In \cite{mattle96}, $50\%$-efficiency coincidence detection of the states $\ket{\Phi^{\pm}}$ was achieved
putting an additional BS  and detector in every
spatial output, which doubles the number of BSs and detectors.  Using the same number of detectors, our analyzer 
allows us to measure in the logical Bell basis with the same detection efficiency for $\ket{\Phi^{\pm}_L}$  as in \cite{mattle96}.
\end{thebibliography}
\end{document}